\documentclass
[twocolumn,aps,amssymb,amsmath,showpacs,superscriptaddress]{revtex4}
\input epsf
\usepackage[dvips]{color}

\begin{document}
\title{
In-medium minijet dissipation in Au+Au collisions at $\sqrt{s_{NN}}$ =
130 and 200~GeV studied with charge-independent two-particle number
fluctuations and correlations.
}
\author
{
Mikhail Kopytine (Kent State University),
for the STAR Collaboration
}
\date{\today}

\begin{abstract}
Medium effects on charged-particle production from minijets are
studied using three complementary analysis techniques.  We find
significant angular collinearity and number correlations on $p_t$ even 
at moderate $p_t<3$ GeV/$c$.  In this $p_t$ range abundant particle
multiplicities enable precision measurements of number correlations of
non-identified hadrons for kinematic variables ($p_t$,$\eta$,$\phi$).
Methods include (1) direct construction of two-particle correlation
functions, (2) inversion of the bin-size dependence of non-statistical
multiplicity fluctuations and (3) two-dimensional discrete wavelet
analysis.

Two-particle correlations on $p_t$ exceed expectations from a
model of equilibrated events with fluctuating global temperature.  A
correlation excess at higher $p_t$ is interpreted as final-state
remnants of initial-state semi-hard collisions.  Lower-$p_t$
correlations exhibit a saddle structure varying strongly with
centrality.  Variations in the forms and relative strengths of low and
high $p_t$ correlations with increasing centrality suggest transport
of semi-hard collision products into the lower $p_t$ region as a
manifestation of in-medium dissipation of minijets.

Correlations on $p_t$ can be associated with angular
correlations on ($\eta$,$\phi$), using analysis methods (1), (2) or
(3).  In particular, wavelet analysis (3) is performed in the
($\eta$,$\phi$) space in bins of $p_t$ ($<2$ GeV/$c$).  Observed
angular correlation structures include those attributed to quantum
correlations and elliptic flow, as well as a localized structure,
increasing in amplitude with $p_t$, and presumed to originate with
minijets.  That structure evolves with increasing centrality in a way
which also suggests dissipation, including an increased correlation
length on $\eta$ which may be related to the influence of a
longitudinally expanding medium on minijet fragmentation.
\end{abstract}
\maketitle 
\twocolumngrid

This note documents experimental observations of medium-modified 
minijet correlation
structures in AuAu collisions at RHIC, presented at a poster session of
the Quark Matter 2004 conference.
The data come from three distinct analyses
which address the same physics topic -- 
quantitative diagnostic of the
strong interaction medium created at RHIC.
The intrinsic properties of this medium \emph{in equilibrium} are connected
with its observable response to the excitations~\cite{FDT}, 
experienced in the course
of heavy ion collision events due to minijet propagation.
In  this work we analyze the response on the basis  of dynamical 
information contained in
fluctuations and correlations in the number density of non-identified hadrons
in the space of kinematic variables $\eta$, $\phi$ and $p_t$.

 A direct construction of two-particle correlations in $p_t$ 
has been performed on 
$\sqrt{s_{NN}}=130$
GeV data.
The basic object of this analysis is density of pairs $\rho$ in the 
two-dimensional
space spanned by the $p_t$ of the two particles.
To make statistical errors more uniform along the kinematic 
variable of choice, 
we replace
$p_t$ by $X(p_t) \equiv 1-\exp[-(m_t-m_{\pi})/0.4 \mbox{GeV}] \in [0,1[$. 
Density of sibling pairs (coming from the same event) $\rho_{sib}$
is compared with a mixed pair reference $\rho_{mix}$.

\begin{figure}[tbh]
\epsfxsize=8.6cm
\epsfbox{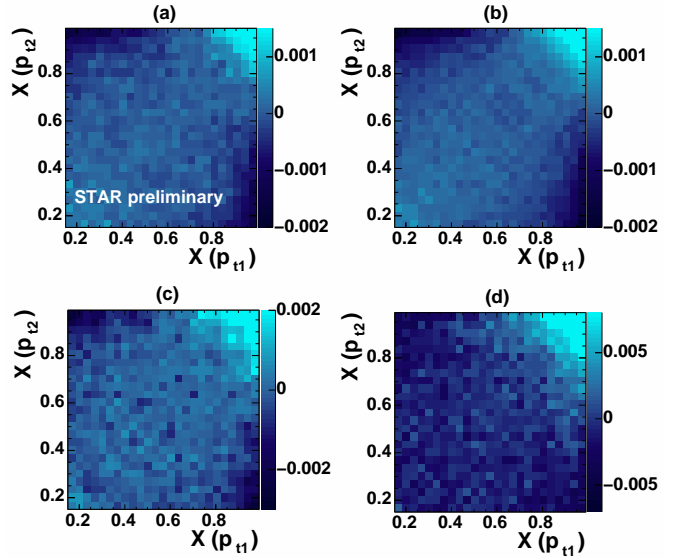}
\caption{\label{Fig:ptxpt_fig2}
$\rho_{sib}/\rho_{mix}[X(p_{t1}),X(p_{t2})]-1$,
pair-density ratios for all unidentified charged primary particles
  for (a) central, (b) mid-central, (c) mid-peripheral, (d) peripheral
AuAu collisions at $\sqrt{s_{NN}}=130 GeV$.
  Note the gray scale differences.
}
\end{figure}

The physics of small momentum differences is known to be dominated by quantum 
statistical
and Coulomb effects.
To focus on the physics of large momentum differences (also referred to as 
large scale),
we eliminate sibling and mixed pairs which simultaneously satisfy conditions
$|\eta_1-\eta_2|<0.3$, $|\phi_1-\phi_2|<\pi/6$,
$|p_{t1}-p_{t2}|<0.15$ GeV/c while having a particle with $p_t<0.8$ GeV/c.
The result in form of a differential density ratio is shown in Fig. 
~\ref{Fig:ptxpt_fig2}
for four centrality classes.
Typical magnitudes of $\rho_{sib}/\rho_{mix}-1$ vary with multiplicity
$\bar{N}$ as $1/\bar{N}$, but there are more subtle shape changes among
the panels of that figure.
To quantify those, we form a fitting model based on a L\'{e}vy distribution
$1/p_t \,dN/\,dp_t \propto [1+\beta_0(m_t-m_\pi)]^{-n}$.
Here the exponent $n$ provides a parameter responsible for ``equilibration'',
$n \rightarrow \infty$ being the Boltzmann limit.
It is informative to use the sum and difference variables 
$m_{t\Sigma} \equiv m_{t1} + m_{t2} - 2m_{\pi}$
and
$m_{t\Delta} \equiv m_{t1} - m_{t2}$;
the ratios in Fig.~\ref{Fig:ptxpt_fig2} are concave along $m_{t\Sigma}$
and convex along $m_{t\Delta}$.
For the mixed pairs, the assumed factorization of two L\'{e}vy distributions
results in
\begin{equation}
\rho_{mix} 
\propto
\Big(1+\frac{\beta_0 m_{t \Sigma}}{2n} \Big)^{-2n}
\Big[
1- \Big(\frac{\beta_0 m_{t \Delta}}{2n+\beta_0 m_{t \Sigma}}\Big)^2 
\Big]^{-n}
\label{Levy_fit}
\end{equation}
For the sibling pairs, the fitting model is as in Eq.~\ref{Levy_fit}, 
except that $n$ in the first and second term is replaced by,
respectively, $n_\Sigma$ and $n_\Delta$.
Curvatures of $\rho_{sib}/\rho_{mix}$ at the origin are
$\beta_0^2( 1/n_\Delta - 1/n )/2$
along $m_{t\Delta}$ and
$\beta_0^2(1/n_\Sigma - 1/n)/2$
along $m_{t\Sigma}$, and have different signs.

The strength of the saddle shape is quantified in terms of the two
curvatures:
\begin{equation}
\Big \{
\frac{\partial^2}{\partial m_{t\Delta}^2} \frac{\rho_{sib}}{\rho_{mix}} <0 ;
\frac{\partial^2}{\partial m_{t\Sigma}^2} \frac{\rho_{sib}}{\rho_{mix}} >0 
\Big \} 
 \Rightarrow  1/n_\Sigma - 1/n_\Delta >0
\end{equation}

\begin{figure}
\epsfxsize=8.6cm
\epsfbox{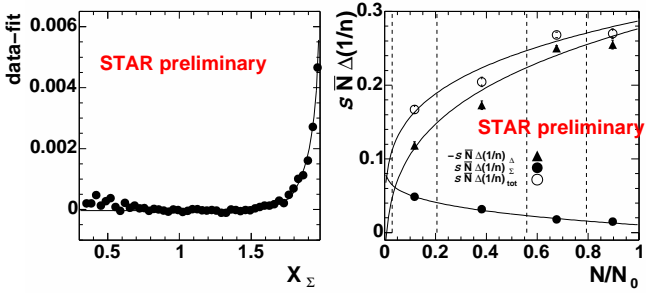}
\caption{
Left: 2D residuals (data$-$''soft'' fit) for central AuAu events
 projected onto $X_\Sigma$.
Shown as a solid curve is the ``hard''
fit based on a hypothesis of a gaussian transverse rapidity distribution.
This hypothesis is consistent with $pp$ data.
Right: centrality trends in the curvature measures.
Curves indicate linear trends on mean path length 
$\nu \approx (N_{part}/2)^{1/3}$.
}
\label{ptxpt_fig4}
\end{figure}

Thus, the correlation structure
is decomposed into \emph{saddle shape} (dissipation)
and \emph{hard component} (high $X(p_t)$ peak). With centrality, 
 the "dissipation" grows with a linear trend on the mean path length,
as seen in  the right panel of Fig.~\ref{ptxpt_fig4}.


Constructing number correlations \emph{directly} 
becomes inefficient when the number of events
and multiplicities are large, since number of computations scales with
the number of particles as $O(N^2)$.
To analyze the $\sqrt{s_{NN}}=200$ GeV data sample, we use 
fluctuation analyses of computational complexity $O(N)$.

In the basis of Fourier harmonics, 
 Wiener-Khinchin theorem relates
autocorrelation  with the local fluctuation power spectrum
via Fourier transform.
Relations between quantities
characterizing fluctuations and correlations
can be also obtained in the basis of box functions (bins).
or in a discrete wavelet basis.

\begin{figure}
\epsfxsize=8.6cm
\epsfbox{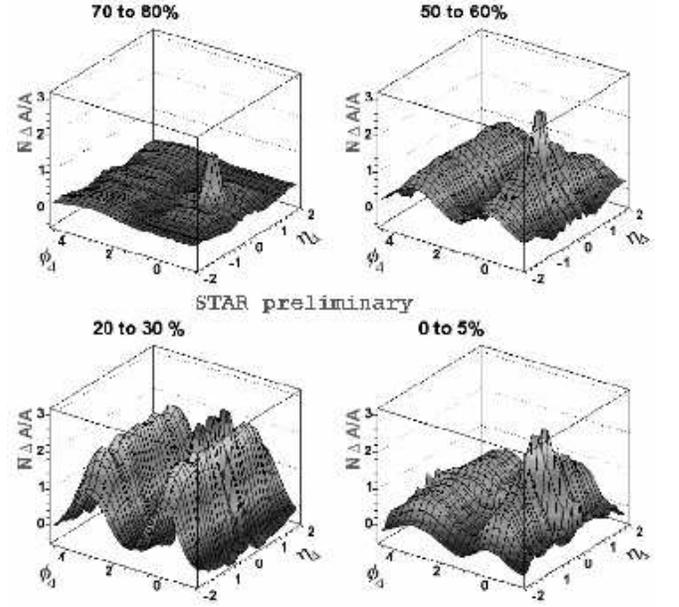}
\caption{Reconstructed autocorrelations in particle number
in AuAu at $\sqrt{s_{NN}}=200$ GeV, for varying centrality.
$\phi_\Delta \equiv \phi_1 - \phi_2$ and
$\eta_\Delta \equiv \eta_1 - \eta_2$.
}
\label{Fig:Duncan_AuAu}
\end{figure}

The inversion of the bin size dependence of the number fluctuations
is performed to express the results of the $O(N)$ 
fluctuation analysis as a two-point autocorrelation (an $O(N^2)$ quantity).
 Relation between 
\emph{number fluctuation} in a $(\delta \eta, \delta \phi)$ bin
and \emph{net autocorrelation} $\Delta A = A_{obj}-A_{ref}$ is
\begin{eqnarray}
\nonumber
\Delta \sigma^2_n(\delta \eta,\delta \phi) = 
\overline{(n-\overline{n})^2}/
\overline{n}-1
\\ =
\int_{0}^{\delta \eta, \delta \phi} \,d\eta_{\Delta} 
 \,d\phi_{\Delta} K(\eta_\Delta, \phi_\Delta)
\frac{\,d^2 \overline{n}}{\,d\eta_\Delta \,d\phi_\Delta}\frac{\Delta A}{A}(\eta_\Delta, \phi_\Delta),
\end{eqnarray}
where $A$ is autocorrelation, $K$ is kernel which depends on
the binning used, $\delta \eta$ and $\delta \phi$ are acceptance ranges,
and $\eta_\Delta$, $\phi_\Delta$ are difference variables.
This integral equation is solved using standard techniques for 
inverse problems to yield the normalized net autocorrelation 
$\frac{\,d^2 \overline{n}}{\,d\eta_\Delta \,d\phi_\Delta}\frac{\Delta A}{A}(\eta_\Delta, \phi_\Delta),$
shown in Fig.~\ref{Fig:Duncan_AuAu}.
Flow $v_2$ structure underlies a same-side minijet peak 
broadened on $\eta_\Delta$ with increasing centrality.

\begin{figure}[tb]
\epsfxsize=8.6cm
\epsfbox{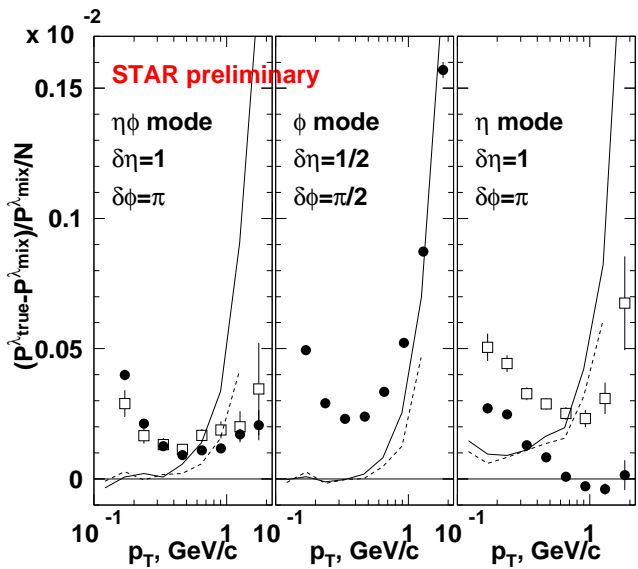}
\caption{
Normalized dynamic texture (see text) as a function of $p_t$ for
central events ($0.6<N/N_0<1.1$).
\textcolor{black}{$\bullet$} -- STAR data;
solid curve -- regular HIJING;
dashed curve -- HIJING with jet quenching;
\textcolor{black}{$\Box$} -- 
peripheral ($N/N_0<0.1$) STAR data, renormalized
according to a "null hypothesis".
}
\label{STAR_central}
\end{figure}

\begin{figure}
\begin{minipage}[t]{55mm}
\epsfxsize=5.5cm
\epsfysize=4.8cm
\epsfbox{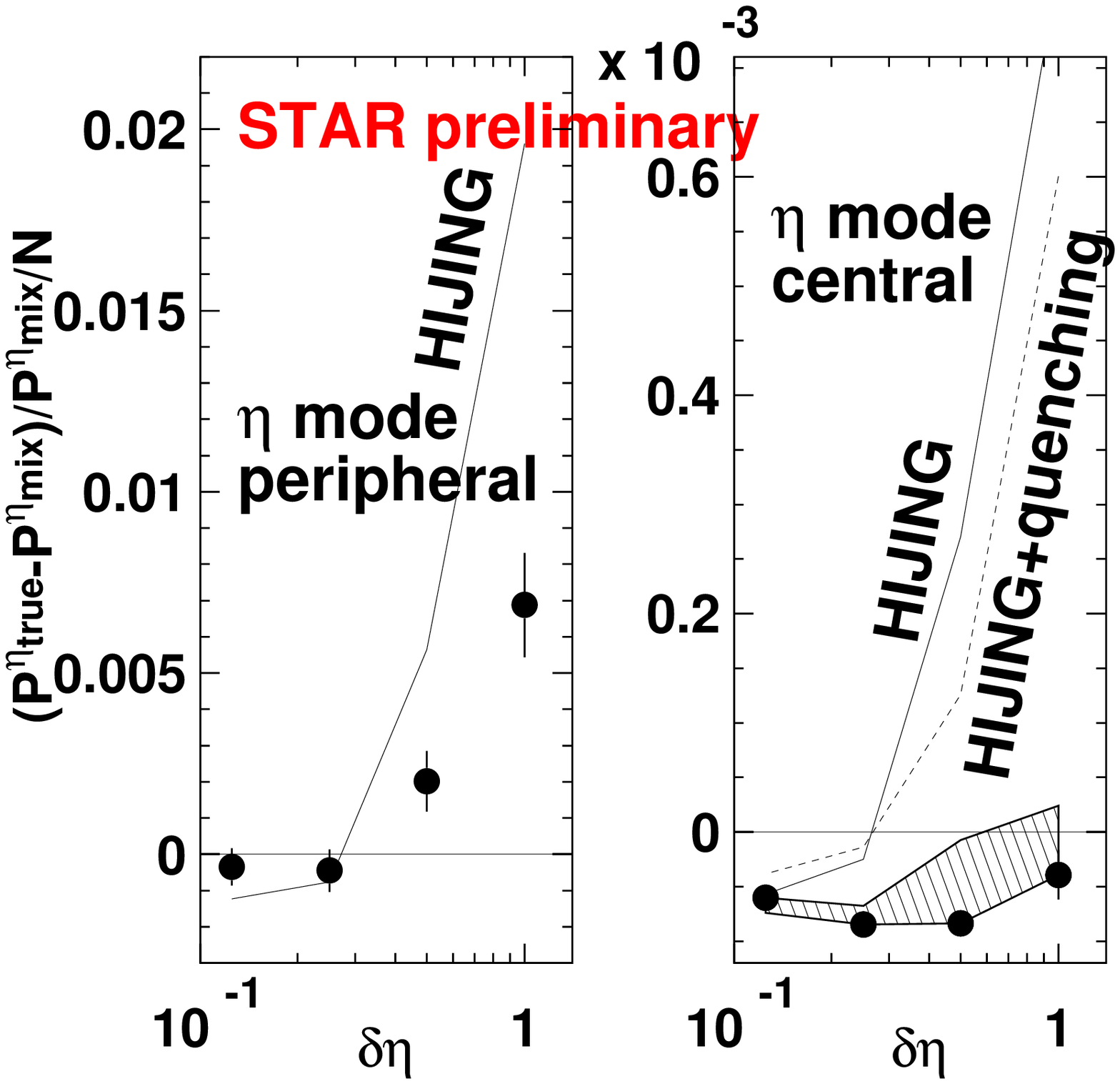}
\end{minipage}
\begin{minipage}[b]{30mm}
\caption{Scale dependence of the dynamic texture (see text) in
peripheral and central events for $1.1<p_t<1.5$ GeV.
\textcolor{black}{$\bullet$} -- STAR;
curves are from HIJING.
An estimate of a systematic error due to track merging
is shown as a hatched area.
}
\label{scale_dependence}
\end{minipage}
\end{figure}

 Application of the DWT power spectrum analysis technique in STAR
is described in detail in ~\cite{DWT_2002}.
The measure of local point-to-point fluctuation is the fluctuation power
$P^\lambda(m)$, constructed out of the DWT expansion coefficients
~\cite{WAILI}
 in the
($\eta,\phi$) space, $a^\lambda_{m,i,j}$, where $\lambda$ indexes the
three possible directional sensitivity modes, pseudorapidity $\eta$,
azimuth $\phi$, and diagonal $\eta\phi$.
\begin{equation}
P^\lambda(m) =
\frac{1}{2^{2m}}\overline{\sum_{i,j}(a^\lambda_{m,i,j})^2} ,
\label{eq:P_m}
\end{equation}
The basic measure of correlation structure is the so-called
\emph{normalized dynamic texture}, 
$(P^\lambda_{true}-P^\lambda_{mix})/P^\lambda_{mix}/N$, 
which incorporates the mixed event reference. 
Here $N$ is a number of particles in an event or subevent
being analyzed.
In case of a one-dimensional random field $X(t)$ such as a
\emph{stationary } time series, the relationship between this $P(m)$
and an autocorrelation is:
\begin{equation}
P(m) = \overline{\int_{-\infty}^{\infty} X(\tau/2)X(-\tau/2) 
W(\tau,m)\,d\tau},
\end{equation}
where $\tau=t_2-t_1$, and $W(\tau,m)$ is the kernel function, localized
and symmetric around 0, which depends only on the wavelet. 
For the Haar wavelet which we use, it is positive around 0 and turns
negative away from 0, therefore one can think of $P(m)$ as a derivative of
an autocorrelation over scale (or $\tau$), averaged over an event sample.

Centrality in the analysis is characterized by the accepted number of
quality tracks in the TPC, $N$, relative to $N_0$, where 
$N_0$ is such that 99\% of minimum bias events have $N<N_0$.

Fig. ~\ref{STAR_central}
shows a difference between dynamic texture data for central
AuAu events and various expectations.
In the experiment, 
we see a change in the $p_t$ 
trend above $p_t=0.6$ GeV in the $\eta$ mode.
Instead of rising with $p_t$ (as in the peripheral events), 
the STAR data points become negative.
This contradicts HIJING~\cite{HIJING}, including the ``quenching'' mode.
In HIJING, the rise of the signal with $p_t$ is obtained by ``turning on'' 
jets.
To make a comparison between central and peripheral data, 
bypassing the model, 
we formulate a ``null hypothesis'': 
the correlation structure 
$(P^\lambda_{true}-P^\lambda_{mix})/P^\lambda_{mix}$ in 
Au+Au collisions 
is invariant of centrality.
Then, the difference in  
$(P^\lambda_{true}-P^\lambda_{mix})/P^\lambda_{mix}/N$ 
between central and peripheral events (including the $p_t$ trends) 
is due to the 
difference in $1/N$ (and in $\,dN/\,dp_t$) 

Shown in Fig. ~\ref{STAR_central} as $\textcolor{black}\Box$ 
is the peripheral data from STAR, 
rescaled under an assumption of the ``null hypothesis'', taking the 
difference in $\,dN/\,dp_t$ into account.
In the $\eta$ mode it is  visible that the actual and rescaled data 
sets differ in both magnitude and $p_t$ trend. 
This underscores the difference of correlation structures between central 
and 
peripheral events and invalidates the ``null hypothesis'' of the $1/N$ 
scaling. 
The left panel shows that jet-like behavior persists almost
unaffected by centrality in  the $\eta\phi$  mode.

We hypothesize that we are observing a \emph{modification} of the minijet 
structure 
predominantly in the longitudinal, $\eta$ direction. 
Longitudinal expansion of the hot and dense medium formed early in the 
collision makes this direction special and is
likely to be part of the modification mechanism. 
If so, we may be observing an effect of the longitudinally expanding 
medium on 
the minijet fragmentation and/or hadronization at ``soft'' $p_t$. 

Fig.~\ref{scale_dependence} 
shows the scale dependence of the correlation structure
in a $p_t$ range where the centrality effect is most pronounced.
We see that the modification of the structure does not imply disappearance
of the correlations in the central events, even on the fine scales.
Thermalization, while affecting the correlations
(more than the popular model~\cite{HIJING} predicts), 
does not result in a correlation-free system in central events, at least
as seen by measuring final state hadrons.

 We conclude 
that large multiplicities of hadrons in the STAR acceptance enable
precision studies of the event correlation structure.
Our observations are consistent with an emerging picture:
minijets from initial state scattering are modified by longitudinally
expanding colored medium.
These measurments of the effect of the medium on the parton fragmentation
and hadronization provide quantitative information
about the medium and nonperturbative QCD.

\end{document}